\documentclass[twocolumn,notitlepage,nofootinbib,superscriptaddress]{revtex4-2}
\usepackage{amsmath,amssymb,bm,graphicx,hyperref}
\allowdisplaybreaks[1]

\renewcommand\d{\partial}

\newcommand\+{\dagger}

\newcommand\D{\mathcal{D}}
\newcommand\E{\mathcal{E}}
\newcommand\J{\mathcal{J}}
\newcommand\M{\mathcal{M}}
\renewcommand\P{\mathcal{P}}
\newcommand\T{\mathcal{T}}
\newcommand\diag{\mathrm{diag}}
\newcommand\eff{\mathrm{eff}}
\newcommand\hydro{\mathrm{hydro}}
\newcommand\CS{\mathrm{CS}}
\newcommand\WZ{\mathrm{WZ}}
\newcommand\g{\mathsf{g}}

\begin{document}

\title{Universal nonlinear responses of quantum Hall systems with Galilean invariance}

\author{Tatsuya Amitani}
\affiliation{Department of Physics, Tokyo Institute of Technology,
Ookayama, Meguro, Tokyo 152-8551, Japan}
\affiliation{Condensed Matter Theory Laboratory, RIKEN, Wako, Saitama 351-0198, Japan}
\author{Yusuke Nishida}
\affiliation{Department of Physics, Tokyo Institute of Technology,
Ookayama, Meguro, Tokyo 152-8551, Japan}

\date{July 2024}

\begin{abstract}
We study two-dimensional systems with Galilean invariance gapped under magnetic fields.
When such quantum Hall systems are coupled with external sources for charge, energy, and momentum currents, they exhibit invariance under the Milne boost as well as under the gauge and general coordinate transformations.
We construct the most general effective action consistent with all the symmetries in the derivative expansion, where an electric field is regarded as order of unity so as to allow for nonlinear responses.
The resulting action is shown to consist of four terms proportional to the Hall conductivity and viscosity and the energy density and magnetization.
We then compute the local currents induced by electromagnetic fields, revealing universal relations among distinct kinds of responses.
In particular, we find the Hall conductivity determining the longitudinal conductivity at nonzero frequency and the Hall viscosity contributing to the nonlinear electrothermal conductivity at nonzero wave number.
\end{abstract}

\maketitle
%\tableofcontents

\section{Introduction}\label{sec:introduction}
Two-dimensional electrons subjected to an external magnetic field exhibit a remarkable phenomenon, the quantum Hall effect~\cite{Klitzing:1980,Tsui:1982}, where the Hall conductivity is quantized as
\begin{align}\label{eq:conductivity}
\sigma_H = \nu\frac{e^2}{2\pi\hbar}.
\end{align}
Here, $\nu$ is an integral or specific rational number corresponding to the filling fraction of Landau levels.
In particular, when electrons are noninteracting, $\nu$ can be expressed as the first Chern number, reflecting the topological structure of single-particle wave functions in momentum space~\cite{Thouless:1982,Kohmoto:1985}.

The quantum Hall systems are equipped with another dissipationless and quantized transport coefficient called the Hall viscosity~\cite{Avron:1995,Avron:1998}.
It is an analog of the Hall conductivity in the viscosity tensor and provided by
\begin{align}\label{eq:viscosity}
\eta_H = \kappa\frac{eB}{8\pi},
\end{align}
where $B$ is the magnetic flux density and $\kappa$ is the Wen-Zee shift multiplied by $\nu$~\cite{Read:2009,Read:2011}.
The latter is an integral or specific rational number relating the number of electrons $N$ to that of magnetic flux quanta $N_\phi$ via $N=\nu N_\phi+(1-\g)\kappa$ when the system is placed on a closed surface with genus $\g$~\cite{Wen:1992}.
For example, $\kappa=\nu^2$ for integer quantum Hall states and $\kappa=1$ for Laughlin states.
Because of its topological nature and universality, the Hall viscosity has recently attracted significant interest across diverse fields in physics~\cite{Hoyos:2014,Berdyugin:2019,Soni:2019}.
In particular, it was shown based on Galilean invariance that the Hall viscosity contributes to the Hall conductivity at nonzero wave number, relating the two distinct responses for charge and momentum~\cite{Hoyos:2012,Bradlyn:2012}.

The purpose of our work is to develop a general framework that enables us to compute charge, energy, and momentum currents of quantum Hall systems induced by electromagnetic fields.
Since the Coulomb interaction between electrons is crucial for the fractional quantum Hall effect, nonperturbative approaches are necessary and we here employ an effective field theory based fully on symmetries.
In particular, previous work along such lines~\cite{Hoyos:2012,Gromov:2014,Geracie:2016} (see also Refs.~\cite{Gromov:2015,Bradlyn:2015} without Galilean invariance) will be extended so as to allow for computation of an energy current and nonlinear responses to full orders in an electric field, which is to reveal universal relations among distinct kinds of responses.

In what follows, we set $\hbar=e=1$.
Greek indices such as $\mu,\nu,\dots$ are valued at $t,x,y$, whereas Latin $a,b,\dots$ and $i,j,\dots$ are valued at $x,y$, and a pair of repeated indices is implicitly summed regardless of their positions.
The totally antisymmetric symbols are denoted by $\epsilon^{\lambda\mu\nu}$ and $\epsilon^{ab}$ with $\epsilon^{txy}=\epsilon^{xy}=1$.
The round (square) brackets enclosing two indices indicate their (anti)symmetrization, such as $C_{(\mu\nu)}\equiv C_{\mu\nu}+C_{\nu\mu}$ and $C_{[\mu\nu]}\equiv C_{\mu\nu}-C_{\nu\mu}$.

\section{Microscopic action and symmetries}
\subsection{Microscopic action and currents}
Let us start with two-dimensional electrons subjected to an external electromagnetic field, whose microscopic action reads $S=S_0+S_\mathrm{int}$ with
\begin{align}\label{eq:flat}
S_0 = \int\!dt\,d^2\!x\left(\Psi^\+i\tensor{D}_t\Psi - \frac{D_i\Psi^\+D_i\Psi}{2m}
+ \frac{gB}{4m}\Psi^\+\Psi\right)
\end{align}
and $\Psi^\+\tensor{D}_\mu\Psi\equiv[\Psi^\+(D_\mu\Psi)-(D_\mu\Psi^\+)\Psi]/2$.
Here, $D_\mu\Psi=(\d_\mu-iA_\mu)\Psi$ and $D_\mu\Psi^\+=(\d_\mu+iA_\mu)\Psi^\+$ are covariant derivatives, and $\phi=-A_t$ and $A_i$ are scalar and vector potentials, respectively.
The Pauli term is added with a $g$ factor for generality, whereas $S_\mathrm{int}$ including interactions between electrons is arbitrary as long as it can be made consistent with local symmetries discussed in Sec.~\ref{sec:symmetry}.

The action coupled with the electromagnetic potential is advantageous to compute an electric current (collectively denoting charge and flux densities) because it can be obtained by differentiating $S$ with respect to $A_\mu$.
Since we are interested also in energy and momentum currents, we wish to couple the action with their external sources.
This is achieved for a relativistic system by placing it on a curved spacetime, where a metric serves as an external source for the energy-momentum tensor.
Similarly, for a nonrelativistic system with Galilean invariance such as ours in Eq.~(\ref{eq:flat}), it should be placed on a curved spacetime called the Newton-Cartan geometry as
\begin{align}\label{eq:curved}
S_0 &= \int\!d^3\!x\sqrt\gamma\,\biggl[v^\mu\Psi^\+i\tensor{D}_\mu\Psi \notag\\
&\quad - \left(\frac{h^{\mu\nu}}{2m}
+ \frac{ig\,\varepsilon^{\lambda\mu\nu}n_\lambda}{4m}\right)D_\mu\Psi^\+D_\nu\Psi\biggr].
\end{align}
Here, the Newton-Cartan metrics are constituted by the clock covector $n_\mu$, the velocity vector $v^\mu$, and the spatial metric $h^{\mu\nu}$ obeying
\begin{align}\label{eq:upper}
n_\mu v^\mu = 1, \qquad n_\mu h^{\mu\nu} = 0.
\end{align}
We also introduce the spatial metric with lower indices according to
\begin{align}\label{eq:lower}
h_{\mu\lambda}h^{\lambda\nu} = P_\mu^\nu \equiv \delta_\mu^\nu - n_\mu v^\nu, \qquad
h_{\mu\nu}v^\nu = 0.
\end{align}
The volume element is then provided by $\sqrt\gamma\,d^3\!x$ with
\begin{align}\label{eq:volume}
\gamma \equiv \det(\gamma_{\mu\nu}), \qquad
\gamma_{\mu\nu} \equiv n_\mu n_\nu + h_{\mu\nu},
\end{align}
where $\gamma_{\mu\nu}$ is invertible by its inverse $v^\mu v^\nu+h^{\mu\nu}$, and the totally antisymmetric tensor reads $\varepsilon^{\lambda\mu\nu}\equiv\epsilon^{\lambda\mu\nu}/\sqrt\gamma$.
The flat spacetime is recovered by setting $n_\mu=v^\mu=(1,0,0)$ and $h^{\mu\nu}=h_{\mu\nu}=\diag(0,1,1)$, where the action on the curved spacetime in Eq.~(\ref{eq:curved}) reduces to the original one in Eq.~(\ref{eq:flat}).
Readers interested in more details about the Newton-Cartan geometry are referred to Refs.~\cite{Son:arxiv,Geracie:2015,Jensen:2018,Jensen:2015} and references therein.

In order to confirm that the Newton-Cartan metrics serve as external sources for energy and momentum currents, we vary the action with respect to the background fields $(A_\mu,n_\mu,v^\mu,h^{\mu\nu})$.
Because of the constraints in Eqs.~(\ref{eq:upper}) and (\ref{eq:lower}), they cannot be varied arbitrarily but the most general variations are $\delta A_\mu$, $\delta n_\mu$,
\begin{align}
\delta v^\mu &= -v^\mu v^\nu\delta n_\nu + P^\mu_\nu\delta\bar v^\nu, \\
\delta h^{\mu\nu} &= -v^{(\mu}h^{\nu)\lambda}\delta n_\lambda
+ P^\mu_\rho P^\nu_\sigma\delta\bar h^{\rho\sigma}, \\
\delta h_{\mu\nu} &= -n_{(\mu}h_{\nu)\lambda}\delta\bar v^\lambda
- h_{\mu\rho}h_{\nu\sigma}\delta\bar h^{\rho\sigma}
\end{align}
parametrized by independent $\delta A_\mu$, $\delta n_\mu$, $\delta\bar v^\mu$, and $\delta\bar h^{\mu\nu}$~\cite{Geracie:2015,Jensen:2018,Jensen:2015}.
We then introduce a current conjugate to each background field according to
\begin{align}\label{eq:variation}
\delta S = \int\!d^3\!x\sqrt\gamma\left(\J^\mu\delta A_\mu - \E^\mu\delta n_\mu
- \P_\mu\delta\bar v^\mu - \frac12\T_{\mu\nu}\delta\bar h^{\mu\nu}\right),
\end{align}
where $\J^\mu$, $\E^\mu$, $\P_\mu$, and $\T_{\mu\nu}$ are to be identified as the electric current, energy current, momentum density, and stress tensor, respectively.
Differentiations of $S_0$ in Eq.~(\ref{eq:curved}) readily lead to
\begin{align}
\J^\mu &= v^\mu\Psi^\+\Psi - \frac{h^{\mu\nu}}{m}\Psi^\+i\tensor{D}_\nu\Psi
+ \frac{g\,\varepsilon^{\lambda\mu\nu}n_\lambda}{4m}\d_\nu(\Psi^\+\Psi), \\\label{eq:current}
\E^\mu &= \frac{v^\mu h^{\nu\lambda} - v^{(\nu}h^{\lambda)\mu}}{2m}D_\nu\Psi^\+D_\lambda\Psi
+ \frac{ig\,\varepsilon^{\mu\nu\lambda}}{4m}D_\nu\Psi^\+D_\lambda\Psi, \\
\P_\mu &= -P_\mu^\nu\Psi^\+i\tensor{D}_\nu\Psi, \\[8pt]
\T_{\mu\nu} &= h_{\mu\nu}\left(v^\lambda\Psi^\+i\tensor{D}_\lambda\Psi
- \frac{h^{\rho\sigma}}{2m}D_\rho\Psi^\+D_\sigma\Psi\right) \notag\\
&\quad + \frac{P_\mu^\rho P_\nu^\sigma}{2m}D_{(\rho}\Psi^\+D_{\sigma)}\Psi,
\end{align}
where transversality of $v^\mu\P_\mu=v^\mu\T_{\mu\nu}=0$ automatically follows, and $\E^\mu$ and $\T_{\mu\nu}$ should suffer additional contributions from $S_\mathrm{int}$.
The resulting expressions in the flat spacetime turn out familiar under the equations of motion for $\Psi$ and $\Psi^\+$ (see Ref.~\cite{Fujii:2018b} for example), confirming the above identifications of currents and the Newton-Cartan metrics serving as their external sources~\cite{currents}.
In particular, $\phi_g$ in $n_\mu=(1+\phi_g,0,0)$ and $v^\mu=[(1+\phi_g)^{-1},0,0]$ corresponds to Luttinger's gravitational potential conjugate to the energy density~\cite{Luttinger:1964}.

\subsection{Symmetries and continuity equations}\label{sec:symmetry}
The nonrelativistic system placed on the Newton-Cartan geometry has another advantage of exhibiting local symmetries (called spurionic symmetries in the high-energy language).
The action in Eq.~(\ref{eq:curved}) is invariant not only under the U(1) gauge transformation,
\begin{align}
\Psi \to e^{i\chi}\Psi, \qquad A_\mu \to A_\mu + \d_\mu\chi,
\end{align}
but also under the general coordinate transformation,
\begin{align}
x^\mu \to x'^\mu = x'^\mu(x),
\end{align}
provided that the background fields with upper (lower) indices are contravariantly (covariantly) transformed.
Furthermore, the defining relations of the Newton-Cartan metrics in Eqs.~(\ref{eq:upper}) and (\ref{eq:lower}) are invariant under locally shifting the velocity field,
\begin{align}\label{eq:milne}
v^\mu &\to v^\mu + h^{\mu\nu}\psi_\nu, \\
h_{\mu\nu} &\to h_{\mu\nu} - n_{(\mu}P_{\nu)}^\lambda\psi_\lambda
+ n_\mu n_\nu h^{\rho\sigma}\psi_\rho\psi_\sigma,
\end{align}
which combined with
\begin{align}\label{eq:gauge}
A_\mu &\to A_\mu + mP_\mu^\nu\psi_\nu
- \frac{m}{2}n_\mu h^{\rho\sigma}\psi_\rho\psi_\sigma \notag\\
&\quad + \frac{g}{4}n_\mu\varepsilon^{\nu\rho\sigma}
\d_\nu(n_\rho P_\sigma^\lambda\psi_\lambda)
\end{align}
keeps the same action invariant~\cite{Jensen:2018,Jensen:2015}.
Such a set of local shifts is known as the Milne boost and may be regarded as a local version of the Galilean boost~\cite{boost}.
We note that $\gamma$ is invariant under the Milne boost in spite of varying $h_{\mu\nu}$.

The invariance of the action under such local transformations implies conservation laws or identities, which can be obtained from $\delta S=0$ with Eq.~(\ref{eq:variation}) assuming the equations of motion for $\Psi$ and $\Psi^\+$.
The gauge invariance then leads to the charge continuity equation,
\begin{align}\label{eq:charge}
\d_\mu\J^\mu = 0,
\end{align}
and the general coordinate invariance to the energy continuity equation,
\begin{align}\label{eq:energy}
\d_\mu\E^\mu = \J^i F_{it},
\end{align}
as well as the momentum continuity equation,
\begin{align}\label{eq:momentum}
\d_t\P_i + \d_j\T_{ij} = \J^\mu F_{i\mu},
\end{align}
where $F_{\mu\nu}=\d_{[\mu}A_{\nu]}$ is the electromagnetic tensor.
On the other hand, the Milne invariance relates the electric current and momentum densities as
\begin{align}\label{eq:identify}
\P_i = m\J^i - \frac{g}{4}\epsilon^{ij}\d_j\J^t,
\end{align}
all presented for the flat spacetime to simplify the expressions.
The corresponding expressions in the curved spacetime can be found in Refs.~\cite{Geracie:2015,Jensen:2018,Jensen:2015}.

Finally, we note that all the discussions above hold including interactions between electrons as long as $S_\mathrm{int}$ coupled with the Newton-Cartan metrics is invariant under the gauge and general coordinate transformations and the Milne boost.
This is typically achieved for an interaction expressed with a local Lagrangian density by introducing auxiliary fields (possibly living in higher dimensions), such as Coulomb, power-law, Yukawa, and any short-range potentials~\cite{Hoyos:2012,Fujii:2018a}.

\section{Effective action and responses}
\subsection{Preliminaries}
In order to compute the expectation values of the charge, energy, and momentum currents induced by electromagnetic fields, we consider the generating functional obtained by integrating out the matter fields,
\begin{align}
e^{iS_\eff} \equiv \int\D\Psi^\+\D\Psi\,e^{iS[\Psi^\+,\Psi;A_\mu,n_\mu,v^\mu,h^{\mu\nu}]}.
\end{align}
Here, $S_\eff=S_\eff[A_\mu,n_\mu,v^\mu,h^{\mu\nu}]$ is the effective action and should be a local functional of the background fields for gapped systems such as quantum Hall systems.
Furthermore, it must inherit all the symmetries from the microscopic action, which together with the derivative expansion constitutes our guiding principle to construct the effective action.
Toward this end, some preparations are necessary.

First, we introduce the vielbein $e^{a\mu}$ according to
\begin{align}\label{eq:vielbein}
h^{\mu\nu} = e^{a\mu}e^{a\nu}, \qquad n_\mu e^{a\mu} = 0,
\end{align}
which rather than the spatial metric will turn out useful to express the effective action in a concise form.
It is not unique, however, because $h^{\mu\nu}$ is invariant under an SO(2) local rotation,
\begin{align}\label{eq:rotation}
e^{x\mu} \pm ie^{y\mu} \to e^{\pm i\theta}(e^{x\mu} \pm ie^{y\mu}),
\end{align}
the invariance under which must also be respected by the effective action.
The spin connection
\begin{align}\label{eq:spin}
\omega_\mu = \frac12\epsilon^{ab}h_{\lambda\nu}e^{a\lambda}\nabla_\mu e^{b\nu}
\end{align}
then acts like an Abelian gauge field under the local rotation, $\omega_\mu\to\omega_\mu+\d_\mu\theta$, where
\begin{align}\label{eq:connection}
\Gamma^\lambda_{\mu\nu} = v^\lambda\d_\mu n_\nu
+ \frac12h^{\lambda\rho}[\d_{(\mu}h_{\nu)\rho} - \d_\rho h_{\mu\nu}]
\end{align}
in $\nabla_\mu e^{a\lambda}=\d_\mu e^{a\lambda}+\Gamma^\lambda_{\mu\nu}e^{a\nu}$ is the torsionful connection for the Newton-Cartan geometry~\cite{Son:arxiv,Geracie:2015}.
Although arbitrary tensors (covectors) can be added to the (spin) connection, such ambiguities do not affect our final result because they can be absorbed into arbitrary coefficients in Eq.~(\ref{eq:action}).

Second, recalling that $v^\mu$, $h_{\mu\nu}$, and $A_\mu$ vary under the Milne boost according to Eqs.~(\ref{eq:milne})--(\ref{eq:gauge}), they are actually inconvenient to construct the effective action.
Therefore, we wish to have a Milne-invariant $u^\mu$ made out of the background fields.
Once such a vector normalized by $n_\mu u^\mu=1$ is found as presented later, the corresponding covector $u_\mu\equiv h_{\mu\nu}u^\nu$ as well as the resulting scalar product $u^2\equiv u_\mu u^\mu$ can be employed to define
\begin{align}
\tilde h_{\mu\nu} &\equiv h_{\mu\nu} - n_{(\mu}u_{\nu)} + n_\mu n_\nu u^2, \\\label{eq:invariant}
\tilde A_\mu &\equiv A_\mu + mu_\mu - \frac{m}{2}n_\mu u^2
+ \frac{g}{4}n_\mu\varepsilon^{\nu\rho\sigma}\d_\nu(n_\rho u_\sigma),
\end{align}
so as to be invariant under the Milne boost~\cite{Jensen:2015}.
The Milne-invariant spin connection $\tilde\omega_\mu$ is then introduced by replacing all $v^\mu$ and $h_{\mu\nu}$ in Eqs.~(\ref{eq:spin}) and (\ref{eq:connection}) with $u^\mu$ and $\tilde h_{\mu\nu}$, respectively.
We note that $\gamma$ remains the same even if $h_{\mu\nu}$ in Eq.~(\ref{eq:volume}) is replaced with $\tilde h_{\mu\nu}$.

Third, we assume that the electromagnetic fields and the Newton-Cartan metrics are order of unity and slowly vary over spacetime, so that derivatives acting on them serve as a small expansion parameter of $O(\d)$ in comparison to gap scales.
Therefore, our power counting scheme in the derivative expansion reads $A_\mu\sim O(\d^{-1})$, $n_\mu,v^\mu,e^{a\mu}\sim O(1)$, and $\d_\mu\sim O(\d)$, the first of which is unproblematic because $A_\mu$ is accompanied by derivatives to ensure the gauge invariance.
In particular, not only the magnetic field but also the electric field is regarded as order of unity so as to allow for nonlinear responses, which go beyond linear responses in previous work based on different frameworks~\cite{Hoyos:2012,Gromov:2014,Geracie:2016}.
With our power counting scheme, the Milne-invariant vector can be constructed order by order, and its leading term at $O(1)$ is uniquely found to be
\begin{align}\label{eq:velocity}
u^\mu = \frac{\varepsilon^{\mu\nu\lambda}\d_\nu A_\lambda}{B} + O(\d),
\qquad B \equiv \varepsilon^{\mu\nu\lambda}n_\mu\d_\nu A_\lambda,
\end{align}
which is the drift velocity invariant under the Milne boost up to $O(\d)$.
Furthermore, $u^\mu$ can be made invariant up to $O(\d^2)$ by adding subleading terms at $O(\d)$, which are not unique but will turn out absent in our effective action.
The drift velocity also appeared in Refs.~\cite{Andreev:2015,Moroz:2015b,Geracie:2016} to serve similar roles.

\subsection{Effective action}
With the above building blocks made out of the background fields, we are ready to construct the most general effective action consistent with all the invariance under the U(1) gauge and general coordinate transformations, the Milne boost, and the SO(2) local rotation.
It is found to consist of the following four terms up to $O(\d)$:
\begin{align}\label{eq:action}
S_\eff &= \int\!d^3\!x\sqrt\gamma\,\biggl[\frac{\nu}{4\pi}
\varepsilon^{\lambda\mu\nu}\tilde A_\lambda\d_\mu\tilde A_\nu
+ \frac{\kappa}{4\pi}\varepsilon^{\lambda\mu\nu}\tilde A_\lambda\d_\mu\tilde\omega_\nu \notag\\
&\quad - \E(\tilde B)
- \M(\tilde B)\varepsilon^{\lambda\mu\nu}n_\lambda\d_\mu n_\nu + O(\d^2)\biggr].
\end{align}
Here, $\nu$ and $\kappa$ are arbitrary constants to ensure the gauge invariance up to surface terms, whereas $\E(\tilde B)$ and $\M(\tilde B)$ are arbitrary functions of $\tilde B\equiv\varepsilon^{\lambda\mu\nu}n_\lambda\d_\mu\tilde A_\nu$.
Their physical meanings implied already by their symbols will be identified later, whose expressions for free fermions are also presented in Appendix~\ref{app:free}.
The first and second terms are known as the Chern-Simons and Wen-Zee actions, respectively, which are modified so as to be invariant under the Milne boost.

The first, second, third, and fourth terms in the effective action are $O(\d^{-1})$, $O(\d)$, $O(1)$, and $O(\d)$, respectively, at their leading orders, whereas each of them is actually mixed order including subleading terms because of such $\tilde A_\mu$ in Eq.~(\ref{eq:invariant}).
Since the Chern-Simons action is primarily $O(\d^{-1})$, the $O(\d)$ correction in Eq.~(\ref{eq:velocity}) denoted by $\delta u^\mu$ may contribute to the effective action up to $O(\d)$.
This is not the case, however, because its contribution is proportional to
\begin{align}
(h_{\lambda\sigma}\delta u^\sigma - n_\lambda h_{\rho\sigma}u^\rho \delta u^\sigma)\,
\varepsilon^{\lambda\mu\nu}\d_\mu A_\nu = 0,
\end{align}
so that $\delta u^\mu\sim O(\d)$ does not enter our effective action.

The Wen-Zee action is concisely expressed with the spin connection, so that its gauge invariance up to a surface term is as manifest as the Chern-Simons action.
Here, the variation of the vielbein constrained by Eq.~(\ref{eq:vielbein}) is provided by
\begin{align}
\delta e^{a\mu} = -v^\mu e^{a\nu}\delta n_\nu + P^\mu_\nu\delta\bar e^{a\nu}
\end{align}
with independent $\delta\bar e^{a\mu}$ related to $\delta\bar h^{\mu\nu}=e^{a(\mu}\delta\bar e^{a\nu)}$.
The stress tensor is then obtained from
\begin{align}\label{eq:stress}
\T_{\mu\nu} = -\frac{h_{\mu\lambda}e^{a\lambda}}{\sqrt\gamma}
\frac{\delta S_\eff}{\delta\bar e^{a\nu}},
\end{align}
which is guaranteed to be symmetric in $\mu\!\leftrightarrow\!\nu$ by the invariance of the effective action under the local rotation in Eq.~(\ref{eq:rotation}).
We note that the Wen-Zee action can also be expressed with the spatial metric as shown in Appendix~\ref{app:wen-zee}, although its gauge invariance up to a surface term is obscured.

\subsection{Universal responses}
It is now straightforward to compute the expectation values of the charge, energy, and momentum currents by differentiating the effective action in Eq.~(\ref{eq:action}) according to Eqs.~(\ref{eq:variation}) and (\ref{eq:stress}).
Because their general expressions are less informative, we present them for the flat spacetime in two physical cases in and out of equilibrium, where electric and magnetic fields are $E_i=\d_{[i}A_{t]}$ and $B=\d_{[x}A_{y]}$, respectively, and the drift velocity reads $u^i=\epsilon^{ij}E_j/B$ from Eq.~(\ref{eq:velocity}).
The resulting local currents are to be completely determined by the four coefficients in the effective action, leading to universal relations among distinct kinds of responses.

When the system is in equilibrium with no electric field but under a static and inhomogeneous magnetic field, we find the electric current to be
\begin{align}
\J^t &= \frac{\nu}{2\pi}B + \d_i\!\left[\left\{m\E''(B) + \frac{\nu g}{8\pi}
- \frac{\kappa}{8\pi}\right\}\frac{\d_iB}{B}\right] + O(\d^3), \\
\J^i &= -\epsilon^{ij}\d_j\E'(B) + O(\d^3),
\end{align}
the energy current to be
\begin{align}
\E^t &= \E(B) + O(\d^2), \\[4pt]
\E^i &= \epsilon^{ij}\d_j\M(B) + O(\d^2),
\end{align}
and the stress tensor to be
\begin{align}
\T_{ij} = \delta_{ij}[B\E'(B) - \E(B)] + O(\d^2)
\end{align}
with the momentum density provided by Eq.~(\ref{eq:identify}).
Here, $\E(B)$ is identified as the energy density and $-\E'(B)$ serves as the charge magnetization with its rotation generating the charge flux density, whereas $\M(B)$ is identified as the energy magnetization with its rotation generating the energy flux density.
The diagonal elements of the stress tensor are $P(B)=B\E'(B)-\E(B)$ identified as the internal pressure~\cite{Cooper:1997,Bradlyn:2012}.

When the system is driven out of equilibrium by a spacetime-dependent electric field under a constant magnetic field, the resulting electric current, energy current, and stress tensor are presented in Appendix~\ref{app:nonlinear} to full orders in the electric field.
Here, we only present its lower-order contributions to highlight our key findings.
First, the stress tensor is found to be
\begin{align}
\T_{ij} &= \delta_{ij}\!\left[B\E'(B) - \E(B)
- \left\{m\E''(B) + \frac{\nu g}{8\pi}\right\}\d_kE_k\right] \notag\\
&\quad + \frac{\kappa}{8\pi}[\d_{(i}E_{j)} - \delta_{ij}\d_kE_k] + O(E^2) + O(\d^2)
\end{align}
up to the linear order in $E_i$.
In particular, the second line in $\T_{ij}$ can be expressed as $\d_{(i}E_{j)}-\delta_{ij}\d_kE_k=-B\epsilon^{(ik}\delta^{j)l}\d_{(k}u_{l)}/2$, which allows us to identify $\kappa B/(8\pi)$ as the Hall viscosity according to Eq.~(\ref{eq:viscosity}).

Then, the electric current is found to be
\begin{align}
\J^t &= \frac{\nu}{2\pi}\left(B - m\frac{\d_iE_i}{B}\right) + O(E^2) + O(\d^3), \\\label{eq:flux}
\J^i &= \frac{\nu}{2\pi}\left(\epsilon^{ij}E_j + m\frac{\d_tE_i}{B}
- m^2\frac{\epsilon^{ij}\d_t^2E_j}{B^2}\right) \notag\\[-2pt]
&\quad + \left[m\E''(B) + \frac{\nu g}{8\pi} - \frac{\kappa}{8\pi}\right]
\frac{\epsilon^{ij}\d_j\d_kE_k}{B} \notag\\[4pt]
&\quad + O(E^2) + O(\d^3)
\end{align}
up to the linear order in $E_i$.
The first term in the first line of $\J^i$ allows us to identify $\nu/(2\pi)$ as the Hall conductivity according to Eq.~(\ref{eq:conductivity}).
On the other hand, the last term in the second line of $\J^i$ shows the Hall viscosity contributing to the Hall conductivity at nonzero wave number, relating the two distinct responses for charge and momentum~\cite{Hoyos:2012,Bradlyn:2012,Son:arxiv,Gromov:2014,Geracie:2015,Geracie:2016}.
Furthermore, the second term in the first line of $\J^i$ indicates that the Hall conductivity also determines the longitudinal conductivity at nonzero frequency, which relates the longitudinal and Hall responses and constitutes one of our new results (see also Appendix~\ref{app:free}).

Finally, the energy current is found to be
\begin{align}
\E^t &= \E(B) - m\E'(B)\frac{\d_iE_i}{B} \notag\\
&\quad + \frac{\nu m}{2\pi}\left(\frac{E^2}{2B}
- m\frac{\epsilon^{ij}E_i\d_tE_j}{B^2}\right) + O(E^3) + O(\d^2), \\
\E^i &= \E'(B)\left(\epsilon^{ij}E_j + m\frac{\d_tE_i}{B}
- \frac{m}{2}\frac{\epsilon^{ij}\d_jE^2}{B^2}\right) \notag\\[-2pt]
&\quad - \left[m\E''(B) + \frac{\nu g}{8\pi}\right]
\frac{\epsilon^{ij}E_j\d_kE_k}{B} \notag\\[2pt]
&\quad - \frac{\kappa}{8\pi}
\frac{\epsilon^{ij}E_k\d_kE_j + \epsilon^{jk}E_j\d_kE_i}{B} + O(E^3) + O(\d^2)
\end{align}
up to the quadratic order in $E_i$.
The nonderivative terms in $\E^\mu$ are consistent with ideal hydrodynamics, where $\E^t|_\hydro=\E(B)+mJ^tu^2/2$ and $\E^i|_\hydro=[P(B)+\E^t]u^i$ hold (see also Appendix~\ref{app:nonlinear}).
Furthermore, the third line in $\E^i$ shows the Hall viscosity contributing also to the nonlinear electrothermal conductivity at nonzero wave number, relating the two distinct responses for energy and momentum, which constitutes the main result of our work.
In particular, the Hall viscosity must enter the charge and energy flux densities consistently in order to satisfy the energy continuity equation in Eq.~(\ref{eq:energy}).

\section{Summary}
In summary, we studied quantum Hall systems with Galilean invariance and their charge, energy, and momentum currents induced by electromagnetic fields with an effective field theory based fully on symmetries.
The resulting local currents are completely determined by the Hall conductivity and viscosity and the energy density and magnetization up to the next-to-next-to-leading orders in the derivative expansion, leading to universal relations among distinct kinds of responses.
In particular, our power counting scheme allows for nonlinear responses to full orders in an electric field, whose formulas are presented in Appendix~\ref{app:nonlinear}.

To highlight our key findings, when an electric field is applied in $x$ direction under a constant magnetic field, the longitudinal conductivity at nonzero frequency is determined by the Hall conductivity according to
\begin{align}
\J^x|_{O(\d E)} = \frac{\sigma_Hm}{B}\d_tE_x,
\end{align}
relating the longitudinal and Hall responses.
Furthermore, the Hall viscosity defined with the stress tensor
\begin{align}
\T_{yy}|_{O(\d E)} = -\left[m\E''(B) + \frac{g}{4}\sigma_H
+ \frac{\eta_H}{B}\right]\d_xE_x
\end{align}
contributes not only to the Hall conductivity at nonzero wave number as
\begin{align}
\J^y|_{O(\d^2E)} = -\left[m\E''(B) + \frac{g}{4}\sigma_H
- \frac{\eta_H}{B}\right]\frac{\d_x^2E_x}{B},
\end{align}
but also to the nonlinear electrothermal conductivity at nonzero wave number as
\begin{align}
\E^y|_{O(\d E^2)} = \left[\frac{m\E'(B)}{B} + m\E''(B) + \frac{g}{4}\sigma_H
+ \frac{\eta_H}{B}\right]\frac{\d_xE_x^2}{2B},
\end{align}
relating the three distinct responses for charge, energy, and momentum.
Such universal relations among distinct kinds of responses of quantum Hall systems with Galilean invariance are worth confirming experimentally and may serve as alternative measurement of the Hall viscosity.

\acknowledgments
The authors thank D.~T.~Son and M.~Watanabe for valuable discussions.
This work was supported by RIKEN Junior Research Associate Program and by JSPS KAKENHI Grant No.\ JP21K03384.

\appendix
\section{Some results for free fermions}\label{app:free}
The effective action in Eq.~(\ref{eq:action}) involves the four coefficients, where $\nu$ and $\kappa$ are independent of interactions because they are quantized to integral or specific rational numbers as discussed in Sec.~\ref{sec:introduction}.
On the other hand, $\E(\tilde B)$ and $\M(\tilde B)$ depend on the interactions between electrons.
Since their functional forms are not universal, we present them only for free fermions in integer quantum Hall states.

Toward this end, we consider a constant magnetic field $B$ with static but inhomogeneous
\begin{align}
n_\mu = (1,n_i), \qquad
h^{\mu\nu} =
\begin{pmatrix}
\delta^{ij}n_in_j & -n_j \\
-n_i & \delta^{ij}
\end{pmatrix}
\end{align}
and trivial $v^\mu=(1,0,0)$, $h_{\mu\nu}=\diag(0,1,1)$, under which the energy density from the effective action reads
\begin{align}\label{eq:density}
n_\mu\E^\mu = \E(B) + 2\M(B)\epsilon^{ij}\d_in_j + O(\d^2n_i).
\end{align}
The microscopic Hamiltonian of free fermions under the same background fields is provided by
\begin{align}
H = \int\!d^2\!x\left[\frac{D_i\Psi^\+D_i\Psi}{2m}
- \frac{gB}{4m}\Psi^\+\Psi + n_i\E^i + O(n_i^2)\right],
\end{align}
where the first two terms are the energy density $\E^t$ and
\begin{align}
\E^i = -\frac{D_{(t}\Psi^\+D_{i)}\Psi}{2m}
- \frac{ig\,\epsilon^{ij}}{4m}D_{[t}\Psi^\+D_{j]}\Psi
\end{align}
is the energy flux density from Eq.~(\ref{eq:current}).
It is then straightforward to show that the first-order perturbation theory in terms of $n_i$ together with the derivative expansion shifts the energy of the $n$th Landau level to
\begin{align}
E_n &= \frac{(2n+1)B}{2m} - \frac{gB}{4m} \notag\\
&\quad - \left[\frac{(2n+1)^2B}{4m^2} - \frac{(2n+1)gB}{4m^2}
+ \frac{g^2B}{16m^2}\right]\epsilon^{ij}\d_in_j \notag\\
&\quad + O(\d^2n_i) + O(n_i^2).
\end{align}
By comparing its summation over $n=0,1,\dots,\nu-1$ with the degeneracy per area $B/(2\pi)$ to Eq.~(\ref{eq:density}), we obtain
\begin{align}
\E(B) &= \left(\frac{\nu^2}{2m} - \frac{\nu g}{4m}\right)\frac{B^2}{2\pi}, \\
\M(B) &= -\left(\frac{4\nu^3-\nu}{12m^2} - \frac{\nu^2g}{4m^2}
+ \frac{\nu g^2}{16m^2}\right)\frac{B^2}{4\pi}
\end{align}
for free fermions in integer quantum Hall states.

Free fermions under a constant magnetic field are also useful to understand our result in Eq.~(\ref{eq:flux}) for a homogeneous but dynamic electric field $E_i$.
The equation of motion for a single electron, $m\dot{u}^i=E_i + \epsilon^{ij}u^jB$, is readily solved by
\begin{align}
u^i = \frac1{B^2+m^2\d_t^2}(B\epsilon^{ij}E_j + m\d_tE_i),
\end{align}
which multiplied by the charge density leads to the charge flux density
\begin{align}
\J^i = \frac{\nu B}{2\pi}u^i
\end{align}
in agreement with the first line of Eq.~(\ref{eq:flux}) under the derivative expansion.
Our result is, however, applicable to interacting electrons as well, being consistent with Kohn's theorem~\cite{Kohn:1961}.

\section{Wen-Zee term in spatial metric}\label{app:wen-zee}
Although the Wen-Zee term is concisely expressed with the spin connection, it can also be expressed with the spatial metric.
To show this, we first define the vielbein with a lower index by $e^a{}_\mu\equiv h_{\mu\nu}e^{a\nu}$ and introduce the torsionless connection
\begin{align}
\mathring\Gamma^\lambda_{\mu\nu} = \frac12v^\lambda\d_{(\mu}n_{\nu)}
+ \frac12h^{\lambda\rho}[\d_{(\mu}h_{\nu)\rho} - \d_\rho h_{\mu\nu}]
\end{align}
in $\mathring\nabla_\mu e^{a\lambda}=\d_\mu e^{a\lambda}+\mathring\Gamma^\lambda_{\mu\nu}e^{a\nu}$.
The spin connection in Eq.~(\ref{eq:spin}) is then expressed as
\begin{align}
\omega_\nu = \frac12\epsilon^{ab}e^a{}_\sigma\mathring\nabla_\nu e^{b\sigma},
\end{align}
whose antisymmetrized derivative leads to
\begin{align}
\d_{[\mu}\omega_{\nu]} &= \mathring\nabla_{[\mu}\omega_{\nu]} \\
&= \frac12\epsilon^{ab}e^a{}_\sigma
\mathring\nabla_{[\mu}\mathring\nabla_{\nu]}e^{b\sigma}
+ \frac12\epsilon^{ab}e^a{}_\sigma
(\mathring\nabla_{[\mu}v^\sigma\mathring\nabla_{\nu]}n_\tau)e^{b\tau} \notag\\
&\quad + \frac12\epsilon^{ab}(\mathring\nabla_{[\mu}e^a{}_\sigma)
P^\sigma_\tau(\mathring\nabla_{\nu]}e^{b\tau}).
\end{align}
By employing $\nabla_\lambda h^{\mu\nu}=0$ and $e^{a\lambda}e^b{}_\lambda=\delta^{ab}$, the last term is found to vanish,
\begin{align}
& \frac12\epsilon^{ab}(\mathring\nabla_{[\mu}e^a{}_\sigma)
P^\sigma_\tau(\mathring\nabla_{\nu]}e^{b\tau}) \notag\\
&= \frac12\epsilon^{ab}(e^{c\sigma}\nabla_{[\mu}e^a{}_\sigma)
(e^{c\tau}\nabla_{\nu]}e^b{}_\tau) = 0,
\end{align}
because $e^{a\lambda}\nabla_\mu e^b{}_\lambda=-e^{b\lambda}\nabla_\mu e^a{}_\lambda$ vanishes for $a=b$.
The Riemann curvature
\begin{align}
{\mathring R^\sigma}{}_{\tau\mu\nu} = \d_{[\mu}\mathring\Gamma^\sigma_{\nu]\tau}
+ \mathring\Gamma^\sigma_{[\mu\rho}\mathring\Gamma^\rho_{\nu]\tau}
\end{align}
results from $\mathring\nabla_{[\mu}\mathring\nabla_{\nu]}e^{b\sigma}={\mathring R^\sigma}{}_{\tau\mu\nu}e^{b\tau}$, which together with $\epsilon^{ab}e^{a\rho}e^{b\tau}=\varepsilon^{\kappa\rho\tau}n_\kappa$ leads to
\begin{align}
\d_{[\mu}\omega_{\nu]} = \frac12\varepsilon^{\kappa\rho\tau}n_\kappa h_{\rho\sigma}
({\mathring R^\sigma}{}_{\tau\mu\nu}
+ \mathring\nabla_{[\mu}v^\sigma\mathring\nabla_{\nu]}n_\tau).
\end{align}
Therefore, the Wen-Zee term turns out to be expressed as
\begin{align}
\varepsilon^{\lambda\mu\nu}A_\lambda\d_\mu\omega_\nu
&= \frac14\varepsilon^{\lambda\mu\nu}\varepsilon^{\kappa\rho\tau}
A_\lambda n_\kappa h_{\rho\sigma} \notag\\
&\quad \times ({\mathring R^\sigma}{}_{\tau\mu\nu}
+ 2\mathring\nabla_\mu v^\sigma\mathring\nabla_\nu n_\tau),
\end{align}
which divided by $A_\lambda$ is a Newton-Cartan analog of the Euler current introduced to describe relativistic quantum Hall systems~\cite{Golkar:2014,Golkar:2015}.
The Euler current is identically conserved and the same expression holds for $u^\mu$, $\tilde h_{\mu\nu}$, and $\tilde A_\mu$ replacing $v^\mu$, $h_{\mu\nu}$, and $A_\mu$, respectively.

\newpage\section{Full nonlinear responses}\label{app:nonlinear}
When we consider the flat spacetime under a spacetime-dependent electric field with a constant magnetic field (i.e., $\d_\mu B=\epsilon^{ij}\d_iE_j=0$), the electric current, the energy current, and the stress tensor to full orders in the electric field are respectively obtained as
\begin{widetext}
\begin{align}
\J^t|_\CS &= \frac{\nu}{2\pi}\left[B - m\frac{\d_iE_i}{B}
+ m^2\d_i\left(\frac{\frac12\d_iE^2 - E_i\d_jE_j}{B^3}\right)\right] + O(\d^3), \\[8pt]
\J^i|_\CS &= \frac{\nu}{2\pi}\left[\epsilon^{ij}E_j
+ m\frac{B\d_tE_i - \frac12\epsilon^{ij}\d_jE^2}{B^2}
+ \frac{g}{4}\frac{\epsilon^{ij}\d_j\d_kE_k}{B}
- m^2\d_t\left(\frac{\epsilon^{ij}B\d_tE_j
+ \frac12\d_iE^2 - E_i\d_jE_j}{B^3}\right)\right. \notag\\
&\quad\left.{} + m^2\epsilon^{ij}\d_j\left(E_k\frac{\epsilon^{kl}B\d_tE_l
+ \frac12\d_kE^2 - E_k\d_lE_l}{B^4}\right)\right] + O(\d^3), \\[8pt]
\E^t|_\CS &= \frac{\nu m}{2\pi}\left(\frac{E^2}{2B} - mE_i
\frac{\epsilon^{ij}B\d_tE_j + \frac12\d_iE^2 - \frac12E_i\d_jE_j}{B^3}\right) + O(\d^2), \\[8pt]
\E^i|_\CS &= \frac{\nu m}{2\pi}\left[\frac{\epsilon^{ij}E_j}{B}\left(\frac{E^2}{2B}
- mE_k\frac{\epsilon^{kl}B\d_tE_l + \frac12\d_kE^2 - E_k\d_lE_l}{B^3}
- \frac{g}{4m}\d_kE_k\right)\right. \notag\\
&\quad\left.{} + \frac{m}{2}E^2
\left(\frac{B\d_tE_i - \frac12\epsilon^{ij}\d_jE^2}{B^4}\right)\right] + O(\d^2), \\[8pt]
\T_{ij}|_\CS &= \frac{\nu m}{2\pi}\left[\frac{\epsilon^{ik}\epsilon^{jl}E_kE_l}{B}
\left(1 + m\frac{\d_nE_n}{B^2}\right)
+ m\epsilon^{(ik}E_k\frac{B\d_tE_{j)} - \frac12\epsilon^{j)l}\d_lE^2}{B^3}
- \frac{g}{4m}\delta_{ij}\d_kE_k\right] + O(\d^2)
\end{align}
from the first term of $S_\eff$,
\end{widetext}
\begin{align}
\J^t|_\WZ &= O(\d^3), \\
\J^i|_\WZ &= -\frac{\kappa}{8\pi}\frac{\epsilon^{ij}\d_j\d_kE_k}{B} + O(\d^3), \\
\E^t|_\WZ &= O(\d^2), \\
\E^i|_\WZ &= -\frac{\kappa}{8\pi}
\frac{\epsilon^{ij}E_k\d_kE_j + \epsilon^{jk}E_j\d_kE_i}{B} + O(\d^2), \\
\T_{ij}|_\WZ &= \frac{\kappa}{8\pi}[\d_{(i}E_{j)} - \delta_{ij}\d_kE_k] + O(\d^2)
\end{align}
from the second term of $S_\eff$,
\begin{align}
\J^t|_\E & = O(\d^3), \\
\J^i|_\E &= \E''(B)m\frac{\epsilon^{ij}\d_j\d_kE_k}{B} + O(\d^3), \\
\E^t|_\E &= \E(B) - \E'(B)m\frac{\d_iE_i}{B} + O(\d^2), \\
\E^i|_\E &= \E'(B)\left(\epsilon^{ij}E_j
+ m\frac{B\d_tE_i - \frac12\epsilon^{ij}\d_jE^2}{B^2}\right) \notag\\
&\quad - \E''(B)m\frac{\epsilon^{ij}E_j\d_kE_k}{B} + O(\d^2), \\[4pt]
\T_{ij}|_\E &= \delta_{ij}[B\E'(B) - \E(B) - \E''(B)m\d_kE_k] + O(\d^2)
\end{align}
from the third term of $S_\eff$, and
\begin{align}
\J^\mu|_\M &= O(\d^3), \\
\E^\mu|_\M &= O(\d^2), \\
\T_{ij}|_\M &= O(\d^2)
\end{align}
from the fourth term of $S_\eff$ with $E^2=E_iE_i$.
The resulting local currents together with the momentum density provided by Eq.~(\ref{eq:identify}) satisfy the continuity equations in Eqs.~(\ref{eq:charge})--(\ref{eq:momentum}).
We note that the nonderivative terms are consistent with ideal hydrodynamics, where $\J^i|_\hydro=\J^tu^i$, $\E^t|_\hydro=\E(B)+mJ^tu^2/2$, $\E^i|_\hydro=[P(B)+\E^t]u^i$, and $\T_{ij}|_\hydro=P(B)\delta_{ij}+mJ^tu^iu^j$ hold (see Ref.~\cite{Fujii:2018b} for example).
Furthermore, $\E^i|_\WZ=\T_{ij}|_\WZ\,u^j$ is consistent with parity-violating hydrodynamics in two dimensions~\cite{Kaminski:2014,Moroz:2015a,Jensen:2015}.


\begin{thebibliography}{99}

\bibitem{Klitzing:1980}
K.~v.~Klitzing, G.~Dorda, and M.~Pepper,
``New method for high-accuracy determination of the fine-structure constant based on quantized Hall resistance,''
\href{https://doi.org/10.1103/PhysRevLett.45.494}
{Phys.\ Rev.\ Lett.\ \textbf{45}, 494-497 (1980)}.

\bibitem{Tsui:1982}
D.~C.~Tsui, H.~L.~Stormer, and A.~C.~Gossard,
``Two-dimensional magnetotransport in the extreme quantum limit,''
\href{https://doi.org/10.1103/PhysRevLett.48.1559}
{Phys.\ Rev.\ Lett.\ \textbf{48}, 1559-1562 (1982)}.

\bibitem{Thouless:1982}
D.~J.~Thouless, M.~Kohmoto, M.~P.~Nightingale, and M.~den~Nijs,
``Quantized Hall conductance in a two-dimensional periodic potential,''
\href{https://doi.org/10.1103/PhysRevLett.49.405}
{Phys.\ Rev.\ Lett.\ \textbf{49}, 405-408 (1982)}.

\bibitem{Kohmoto:1985}
M.~Kohmoto,
``Topological invariant and the quantization of the Hall conductance,''
\href{https://doi.org/10.1016/0003-4916(85)90148-4}
{Ann.\ Phys.\ (NY) \textbf{160}, 343-354 (1985)}.

\bibitem{Avron:1995}
J.~E.~Avron, R.~Seiler, and P.~G.~Zograf,
``Viscosity of quantum Hall fluids,''
\href{https://doi.org/10.1103/PhysRevLett.75.697}
{Phys.\ Rev.\ Lett.\ \textbf{75}, 697-700 (1995)}.
%arXiv:cond-mat/9502011

\bibitem{Avron:1998}
J.~E.~Avron,
``Odd viscosity,''
\href{https://doi.org/10.1023/A:1023084404080}
{J.\ Stat.\ Phys.\ \textbf{92}, 543-557 (1998)}.
%arXiv:physics/9712050 [physics.flu-dyn]

\bibitem{Read:2009}
N.~Read,
``Non-Abelian adiabatic statistics and Hall viscosity in quantum Hall states and $p_x+ip_y$ paired superfluids,''
\href{https://doi.org/10.1103/PhysRevB.79.045308}
{Phys.\ Rev.\ B \textbf{79}, 045308 (2009)}.
%arXiv:0805.2507 [cond-mat.mes-hall]

\bibitem{Read:2011}
N.~Read and E.~H.~Rezayi,
``Hall viscosity, orbital spin, and geometry: Paired superfluids and quantum Hall systems,''
\href{https://doi.org/10.1103/PhysRevB.84.085316}
{Phys.\ Rev.\ B \textbf{84}, 085316 (2011)}.
%arXiv:1008.0210 [cond-mat.mes-hall]

\bibitem{Wen:1992}
X.~G.~Wen and A.~Zee,
``Shift and spin vector: New topological quantum numbers for the Hall fluids,''
\href{https://doi.org/10.1103/PhysRevLett.69.953}
{Phys.\ Rev.\ Lett.\ \textbf{69}, 953-956 (1992)}.

\bibitem{Hoyos:2014}
C.~Hoyos,
``Hall viscosity, topological states and effective theories,''
\href{https://doi.org/10.1142/S0217979214300072}
{Int.\ J.\ Mod.\ Phys.\ B \textbf{28}, 1430007 (2014)}.
%arXiv:1403.4739 [cond-mat.mes-hall]

\bibitem{Berdyugin:2019}
A.~I.~Berdyugin, S.~G.~Xu, F.~M.~D.~Pellegrino, R.~Krishna Kumar, A.~Principi, I.~Torre, M.~Ben Shalom, T.~Taniguchi, K.~Watanabe, I.~V.~Grigorieva, M.~Polini, A.~K.~Geim, and D.~A.~Bandurin,
``Measuring Hall viscosity of graphene's electron fluid,''
\href{https://doi.org/10.1126/science.aau0685}
{Science \textbf{364}, 162-165 (2019)}.
%arXiv:1806.01606 [cond-mat.mes-hall]

\bibitem{Soni:2019}
V.~Soni, E.~S.~Bililign, S.~Magkiriadou, S.~Sacanna, D.~Bartolo, M.~J.~Shelley, and W.~T.~M.~Irvine,
``The odd free surface flows of a colloidal chiral fluid,''
\href{https://doi.org/10.1038/s41567-019-0603-8}
{Nat.\ Phys.\ \textbf{15}, 1188-1194 (2019)}.
%arXiv:1812.09990 [physics.flu-dyn]

\bibitem{Hoyos:2012}
C.~Hoyos and D.~T.~Son,
``Hall viscosity and electromagnetic response,''
\href{https://doi.org/10.1103/PhysRevLett.108.066805}
{Phys.\ Rev.\ Lett.\ \textbf{108}, 066805 (2012)}.
%arXiv:1109.2651 [cond-mat.mes-hall]

\bibitem{Bradlyn:2012}
B.~Bradlyn, M.~Goldstein, and N.~Read,
``Kubo formulas for viscosity: Hall viscosity, Ward identities, and the relation with conductivity,''
\href{https://doi.org/10.1103/PhysRevB.86.245309}
{Phys.\ Rev.\ B \textbf{86}, 245309 (2012)}.
%arXiv:1207.7021 [cond-mat.stat-mech]

\bibitem{Gromov:2014}
A.~Gromov and A.~G.~Abanov,
``Density-curvature response and gravitational anomaly,''
\href{https://doi.org/10.1103/PhysRevLett.113.266802}
{Phys.\ Rev.\ Lett.\ \textbf{113}, 266802 (2014)}
%arXiv:1403.5809 [cond-mat.str-el]

\bibitem{Geracie:2016}
M.~Geracie, K.~Prabhu, and M.~M.~Roberts,
``Covariant effective action for a Galilean invariant quantum Hall system,''
\href{https://doi.org/10.1007/JHEP09(2016)092}
{J.\ High Energy Phys.\ 09 (2016) 092}.
%arXiv:1603.08934 [cond-mat.mes-hall]

\bibitem{Gromov:2015}
A.~Gromov and A.~G.~Abanov,
``Thermal Hall effect and geometry with torsion,''
\href{https://doi.org/10.1103/PhysRevLett.114.016802}
{Phys.\ Rev.\ Lett.\ \textbf{114}, 016802 (2015)}.
%arXiv:1407.2908 [cond-mat.str-el]

\bibitem{Bradlyn:2015}
B.~Bradlyn and N.~Read,
``Low-energy effective theory in the bulk for transport in a topological phase,''
\href{https://doi.org/10.1103/PhysRevB.91.125303}
{Phys.\ Rev.\ B \textbf{91}, 125303 (2015)}.
%arXiv:1407.2911 [cond-mat.mes-hall]

\bibitem{Son:arxiv}
D.~T.~Son,
``Newton-Cartan geometry and the quantum Hall effect,''
\href{https://doi.org/10.48550/arXiv.1306.0638}
{arXiv:1306.0638 [cond-mat.mes-hall]}.

\bibitem{Geracie:2015}
M.~Geracie, D.~T.~Son, C.~Wu, and S.-F.~Wu,
``Spacetime symmetries of the quantum Hall effect,''
\href{https://doi.org/10.1103/PhysRevD.91.045030}
{Phys.\ Rev.\ D \textbf{91}, 045030 (2015)}.
%arXiv:1407.1252 [cond-mat.mes-hall]

\bibitem{Jensen:2018}
K.~Jensen,
``On the coupling of Galilean-invariant field theories to curved spacetime,''
\href{https://doi.org/10.21468/SciPostPhys.5.1.011}
{SciPost Phys.\ \textbf{5}, 011 (2018)}.
%arXiv:1408.6855 [hep-th]

\bibitem{Jensen:2015}
K.~Jensen,
``Aspects of hot Galilean field theory,''
\href{https://doi.org/10.1007/JHEP04(2015)123}
{J.\ High Energy Phys.\ 04 (2015) 123}.
%arXiv:1411.7024 [hep-th]

\bibitem{Fujii:2018b}
K.~Fujii and Y.~Nishida,
``Hydrodynamics with spacetime-dependent scattering length,''
\href{https://doi.org/10.1103/PhysRevA.98.063634}
{Phys.\ Rev.\ A \textbf{98}, 063634 (2018)}.
%arXiv:1807.07983 [cond-mat.quant-gas]

\bibitem{currents}
Although our covariant currents differ from the noncovariant currents in Ref.~\cite{Geracie:2015} and the Milne-invariant currents in Ref.~\cite{Jensen:2018}, all of them coincide in the flat spacetime where our results are to be presented.

\bibitem{Luttinger:1964}
J.~M.~Luttinger,
``Theory of thermal transport coefficients,''
\href{https://doi.org/10.1103/PhysRev.135.A1505}
{Phys.\ Rev.\ \textbf{135}, A1505-A1514 (1964)}.

\bibitem{boost}
The Galilean boost corresponds to the special case of constant $\psi_i=V^i$ together with $x'^i-x^i=-V^it$ and $\chi=mV^2t/2-mV^ix^i$ in the flat spacetime.

\bibitem{Fujii:2018a}
K.~Fujii and Y.~Nishida,
``Low-energy effective field theory of superfluid $^3$He-B and its gyromagnetic and Hall responses,''
\href{https://doi.org/10.1016/j.aop.2018.06.003}
{Ann.\ Phys.\ (NY) \textbf{395}, 170-182 (2018)}.
%arXiv:1610.06330 [cond-mat.supr-con]

\bibitem{Andreev:2015}
O.~Andreev,
``More on nonrelativistic diffeomorphism invariance,''
\href{https://doi.org/10.1103/PhysRevD.91.024035}
{Phys.\ Rev.\ D \textbf{91}, 024035 (2015)}.
%arXiv:1408.7031 [hep-th]

\bibitem{Moroz:2015b}
S.~Moroz, C.~Hoyos, and L.~Radzihovsky,
``Galilean invariance at quantum Hall edge,''
\href{https://doi.org/10.1103/PhysRevB.91.195409}
{Phys.\ Rev.\ B \textbf{91}, 195409 (2015)}.
%arXiv:1502.00667 [cond-mat.str-el]

\bibitem{Cooper:1997}
N.~R.~Cooper, B.~I.~Halperin, and I.~M.~Ruzin,
``Thermoelectric response of an interacting two-dimensional electron gas in a quantizing magnetic field,''
\href{https://doi.org/10.1103/PhysRevB.55.2344}
{Phys.\ Rev.\ B \textbf{55}, 2344-2359 (1997)}.
%arXiv:cond-mat/9607001

\bibitem{Kohn:1961}
W.~Kohn,
``Cyclotron resonance and de Haas-van Alphen oscillations of an interacting electron gas,''
\href{https://doi.org/10.1103/PhysRev.123.1242}
{Phys.\ Rev.\ \textbf{123}, 1242-1244 (1961)}.

\bibitem{Golkar:2014}
S.~Golkar, M.~M.~Roberts, and D.~T.~Son,
``Effective field theory of relativistic quantum Hall systems,''
\href{https://doi.org/10.1007/JHEP12(2014)138}
{J.\ High Energy Phys.\ 12 (2014) 138}.
%arXiv:1403.4279 [cond-mat.mes-hall]

\bibitem{Golkar:2015}
S.~Golkar, M.~M.~Roberts, and D.~T.~Son,
``The Euler current and relativistic parity odd transport,''
\href{https://doi.org/10.1007/JHEP04(2015)110}
{J.\ High Energy Phys.\ 04 (2015) 110}.
%arXiv:1407.7540 [hep-th]

\bibitem{Kaminski:2014}
M.~Kaminski and S.~Moroz,
``Nonrelativistic parity-violating hydrodynamics in two spatial dimensions,''
\href{https://doi.org/10.1103/PhysRevB.89.115418}
{Phys.\ Rev.\ B \textbf{89}, 115418 (2014)}.
%arXiv:1310.8305 [cond-mat.mes-hall]

\bibitem{Moroz:2015a}
S.~Moroz and C.~Hoyos,
``Effective theory of two-dimensional chiral superfluids: Gauge duality and Newton-Cartan formulation,''
\href{https://doi.org/10.1103/PhysRevB.91.064508}
{Phys.\ Rev.\ B \textbf{91}, 064508 (2015)}.
%arXiv:1408.5911 [cond-mat.quant-gas]

\end{thebibliography}
\end{document}